# Decoupled measurement and modeling of interface reaction kinetics of ion-intercalation battery electrodes

Ruoyu Xiong [a], Mengyuan Zhou [a], Longhui Li [a], Jia Xu [a], Maoyuan Li [a], Bo yan [b], Dequn Li [a], Yun Zhang [a, *], Huamin Zhou [a, **]

* Corresponding author: marblezy@hust.edu.cn, Tel: 86-27-87543492

** Corresponding author: hmzhou@hust.edu.cn, Tel: 86-27-87543492

a. State Key Laboratory of Material Processing and Die & Mould Technology, School of Materials Science and Engineering, Huazhong University of Science and Technology, Wuhan 430074, China

b. School of Materials Science and Engineering, Shanghai Jiao Tong University, Shanghai, 200030, China

## Abstract

Ultrahigh rate performance of active particles used in lithium-ion battery electrodes has been revealed by single-particle measurements, which indicates a huge potential for developing high-power batteries. However, the charging/discharging behaviors of single particles at ultrahigh C-rates can no longer be described by the traditional electrochemical kinetics in such ion-intercalation active materials. In the meantime, regular kinetic measuring methods meet a challenge due to the coupling of interface reaction and solid-state diffusion processes of active particles. Here, we decouple the reaction and diffusion kinetics via time-resolved potential measurements with an interval of 1 ms, revealing that the classical Butler-Volmer equation deviates from the actual relation between current density, overpotential, and $Li^+$ concentration. An interface ion-intercalation model is developed which considers the excess driving force of $Li^+$ (de)intercalation in the charge transfer reaction for ion-intercalation materials. Simulations demonstrate that the proposed model enables accurate prediction of charging/discharging at both single-particle and electrode scales for various active materials. The kinetic limitation processes from single particles to composite electrodes are systematically revealed, promoting rational designs of high-power batteries.

Keywords: High-power; Single particles; Ultrahigh rate; Ion-intercalation model; Potential measurements

# 1 Introduction

Battery technologies entered a phase of rapid development since lithium-ion batteries (LIBs) were commercialized, and at the same time, emerging markets continually put forward higher requirements for battery performances. More excellent rate performance (capacity retention ability in fast charging/discharging) is demanded in many high-power applications [1-3], including electric vehicles (e.g. fast charging and jump starters), electric power tools, unmanned aerial vehicles, mobile robots, and electric aircraft. Electrode active particles are believed to determine the upper limit of battery rate performance [4-9], which are ion-intercalation compounds and serve as the elementary units of electrochemical energy storage in LIBs. For instance, significantly enhanced interface reaction kinetics and solid-state diffusion were achieved via modulating the surface/interface [8,10] and crystallography of active particles [5], respectively. Thus, understanding the kinetic behaviors of active particles is the foundation of developing high-power LIBs.

However, it is difficult to acquire accurate kinetic data of active particles by testing a cell due to the decoupling effects of complicated kinetic processes and the effect of composite electrode microstructures [8,11]. For example, the reported values of the exchange current density and Li diffusion coefficient vary by several orders of magnitude for the same material [12-14]. To address this issue, Uchida, Kanamura et al. developed single-particle measurement technology to decouple the kinetic processes of active particles from the entire electrode [15-17]. Impressive ultrahigh rate performance found in various active particles [11,12,18-22] exposed the invalidation of conventional model prediction at the single-particle scale [23-26]. Besides, interfacial reaction and solid-state diffusion within the active particles cannot be further decoupled by the charging/discharging tests, leading to a debate whether the practical Tafel plots of such ion-intercalation active electrodes obey the classical Butler-Volmer equation in electrochemical kinetics [27-29]. Therefore, there is a critical need for decoupled kinetic measurements of active particles and reconsideration of the interface reaction model.

In this work, time-resolved potential measurements with an interval of 1ms were applied to decouple the interface reaction and solid-state diffusion processes of single active particles in terms of their distinct time characteristics, deriving separated interface reaction overpotentials for model verification. An interface ion-intercalation model was proposed by considering the Li$^+$ (de)intercalation driving force in ion-intercalation active electrodes. The developed model was further validated by simulating the ultrahigh-rate charging/discharging of various reported active materials in the single-particle measurements. Moreover, the model validity was also demonstrated in predicting composite electrode performance, revealing different kinetic limiting processes with varied C-rates and areal capacity for rational designs of high-power batteries.

## 2 Interface reaction kinetics beyond the Butler-Volmer equation

LIB electrode active particles are intercalation compounds which allow ion intercalation at the electrode/electrolyte interface and in the bulk phase (also termed solid-state diffusion) during charging/discharging. Regular sampling speeds (1 s - 1 min) of voltage/potential signals are unable to capture the dynamic behaviors of fast interface reaction at a scale of milliseconds [30-33]. To this end, time-resolved potential measurements with an interval of 1 ms were performed on a single-layer particle electrode (SLPE) to separate the potential responses of interface reaction and solid-state diffusion to a step current stimulation (**Fig. 1**a, the detailed methodology can be found in Ref. [34]. The interface reaction overpotentials ($\eta$) hence can be obtained under different current densities and surface Li$^+$ concentration of active particles (**Fig. 1**b,c), compared with the Butler-Volmer kinetic model which is commonly applied to describe the current-overpotential relation at electrode/electrolyte interfaces [35-37] (**Eqn. 1**).

$$i_a = i_0 \left[ \exp\left(\frac{\alpha F}{RT}\eta_{ct}\right) - \exp\left(-\frac{(1-\alpha)F}{RT}\eta_{ct}\right) \right] \qquad (1)$$

where $i_0$ is the exchange current density in connection with the concentrations of active sites/vacancies ($c_{site}$), solid-state Li$^+$ ($c_s$), and liquid-state Li$^+$ ($c_l$) [38,39]:

$$i_0 = Fkc_{site}^{\alpha} c_s^{1-\alpha} c_l^{\alpha} \qquad (2)$$

It is found that the practical relation between current density $i_a$ and overpotential $\eta$ do not obey this model at ultrahigh C-rates, as shown in **Fig. 1**b. At large overpotentials, the Butler-Volmer model is known to be simplified to the Tafel equation (**Eqn. 3**), which is a straight line with a slope of $\frac{\alpha F}{2.303RT}$ (Tafel slope) in logarithmic coordinates.

$$\log i_a = \log i_0 + \frac{\alpha F}{2.303RT}\eta \qquad (3)$$

In contrast, the experimental results show a curved Tafel plot when the interface reaction overpotential is over 100 mV. Similar phenomena were also found in the single-particle measurement results of various active materials [11,12,18-22,27], where the curved Tafel plots were attributed to solid concentration overpotential since it had not been eliminated from the measured total overpotential. Nevertheless, the concentration overpotential caused by solid-state Li$^+$ diffusion is much smaller than the excess overpotential (the part that exceeds the value calculated by the Butler-Volmer model); for instance, the whole range of potential plateau of LiCoO$_2$ (LCO) is about 0.3 V while the excess overpotential can reach 0.6 V at the beginning of discharging [19]. More importantly, our decoupled measurements here directly demonstrate that the curved Tafel plots inherently exist in such ion-intercalation active materials. **Fig. 1**c shows the dependence of overpotential $\eta$ on solid-state Li$^+$ concentration ($x = c_s/c_{s,max}$) of active particle surfaces, which also deviates from the Butler-Volmer model at a high level of Li$^+$ concentration ($x > 0.8$). Therefore, we speculated that the excess overpotential originates from the concentration-dependent Li$^+$ insertion/extraction step in the solid-phase side of the electrode/electrolyte interface, instead of other steps occurring on the liquid-phase side (e.g. ion desolvation).

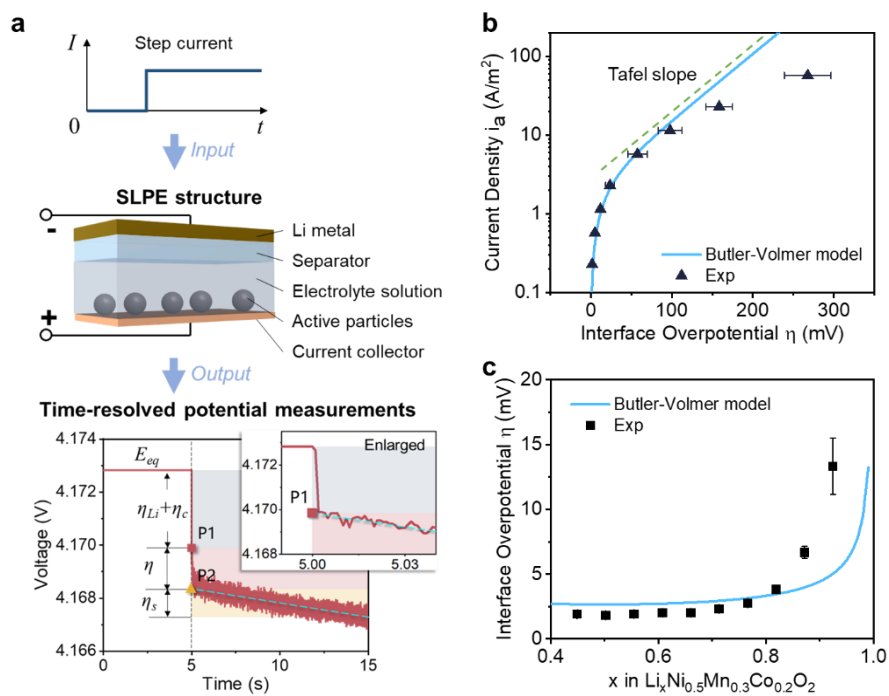

**Fig. 1 Decoupled measurements of interface reaction overpotential.** a. schematic of the overpotential decomposition technology. Time-resolved potential measurements with an interval of 1 ms were performed on SLPE-based cells to separate the overpotentials of Li metal anode ($\eta_{Li}$), current collector-active particle contact ($\eta_c$), interface reaction ($\eta$) and solid-state diffusion ($\eta_s$) of active particles. b. Current density-overpotential relation of interface reaction compared with the Butler-Volmer model. c. Dependence of interface reaction overpotential on solid-state Li$^+$ concentration of active particle surfaces.

## 3 Development of interface ion-intercalation model

As is known, the Butler-Volmer model was developed based on the phenomenology of traditional electrochemical reactions with inert electrodes and solvated ions in electrolyte solutions [28] (**Fig. 2**a). However, it may be not appropriate for solid-liquid reactions with active materials, for example, ion-intercalation compounds, which have been used as electrode materials of rechargeable batteries since the 1970s. As illustrated in **Fig. 2**b, in ion-intercalation reactions, the activation energy barrier of the reactant ions or active sites in the solid state is supposed to be higher than that in the liquid state due to the restraint of the host material, which means it needs more driving force (overpotential) to break the bonds between atoms/ions. Besides, with the increase of solid-state $Li^+$ concentration, the transformation of lattice structure leads to a stronger interaction between the intercalated ions and the host material, which manifests in an incremental overpotential (**Fig. 1**c). Such correlation between the current density, overpotential, and $Li^+$ concentration at the particle/electrolyte interface may not be accurately described by the Butler-Volmer model which is based on an exponential form proposed by Arrhenius. To this end, the overpotential of $Li^+$ (de)intercalation is considered combined with the part of charge transfer in the interface reaction process, deriving the total interface overpotential as

$$\eta = \eta_{ct} + \eta_{in} \tag{4}$$

where $\eta_{ct}$ denotes the part of charge transfer, and $\eta_{in}$ is the part of. $Li^+$ (de)intercalation. Assuming that the resistance of $Li^+$ (de)intercalation at the surface of the host material can be treated as a film resistance of conductors, then $\eta_{in}$ can be expressed as

$$\frac{\eta_{in}}{i_a} = R_{in} = \frac{d}{\sigma(c_s, c_{site})A} \tag{5}$$

where $d$ denotes the hypothetical thickness of the first lattice layer of the host material for $Li^+$ intercalation (**Fig. 2**b), $A$ the surface/interface area, and $\sigma$ the equivalent ionic conductivity which is related to the concentrations of sites/$Li^+$ at the first lattice layer. According to the unimodal conductive property of concentrated solutions [40,41], the equivalent ionic conductivity $\sigma$ is set to be determined by the $Li^+$ concentration at a lower level of $Li^+$ concentration and by the concentration of remain active sites at a high level of $Li^+$ concentration. With a linear assumption that $\sigma$ is approximately proportional to the accessible site/$Li^+$ concentration at the material surface in the

corresponding range of Li$^+$ concentration, we obtain the piecewise function of $\sigma(c_{Li}, c_{site})$

$$\sigma(c_{Li}, c_{site}) = \begin{cases} \sigma_0 \dfrac{c_s}{c_{s,\max}} &, \quad 0 < x < x_c \\ \sigma_0 \dfrac{c_{site}}{c_{s,\max}} &, \quad x_c \leq x < 1 \end{cases} \tag{6}$$

where the theoretical maximum concentration provided for Li$^+$ intercalation of the host material $c_{s,\max} > c_s + c_{site}$ in practice, and the critical Li$^+$ concentration $x_c$ is related to the lattice property of the host material. With the charge transfer overpotential ($\eta_{ct}$) calculated by the Butler-Volmer model, the total interface overpotential can be given by substituting **Eqn. 5** into **Eqn. 4**

$$\eta = \frac{2RT}{F} \operatorname{arcsinh}\left(\frac{i_a}{2i_0}\right) + \frac{\rho c_{s,\max}}{\sigma(c_s, c_{site})} i_a \tag{7}$$

where $\rho = d/(\sigma_0 A)$. **Eqn. 7** is the interface ion-intercalation model for the electrochemical kinetics of ion-intercalation active materials.

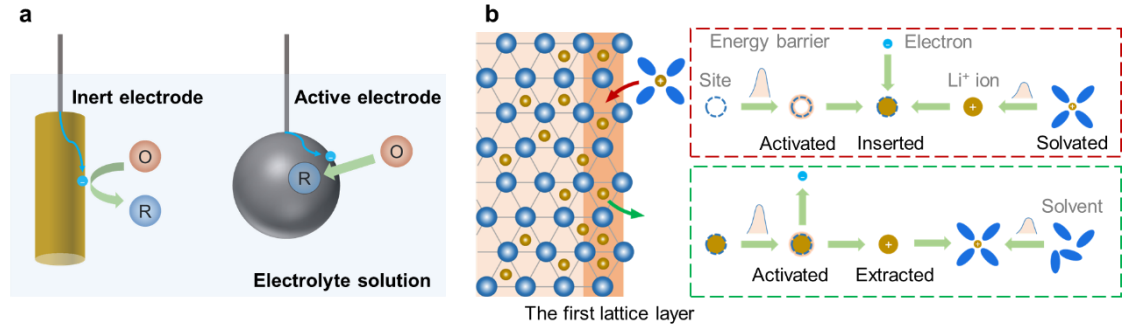

**Fig. 2 Schematic of interface reaction kinetics of ion-intercalation active electrodes.** a. Comparison of interface electrochemical reactions on inert electrodes and active electrodes. O: oxidation state; R: reduction state. b. Physical picture of the interface ion-intercalation model for ion-intercalation active electrodes.

## 4 Validation and applications

### 4.1 Validating the interface ion-intercalation model

**Fig. 3** demonstrates that the interface ion-intercalation model well describes the correlation between the current density, overpotential, and Li$^+$ concentration in reaction kinetics of different ion-intercalation materials. The estimated model parameters are listed in **Table 1**, including the charge transfer constant $k$ and the Li$^+$ (de)intercalation constant $\rho$. The charge transfer constants $k$ of typical layered transition metal oxides, LiNi$_{0.5}$Mn$_{0.3}$Co$_{0.2}$O$_2$ (NMC532) and LCO, are both in the order of $10^{-11}$ m$^{2.5}$ mol$^{-0.5}$ s$^{-1}$, one order higher than that of natural graphite material ($10^{-12}$ m$^{2.5}$ mol$^{-0.5}$ s$^{-1}$). Whereas, the Li$^+$ (de)intercalation constants $\rho$ of NMC532 and LCO ($10^{-3}$ Ω m$^2$) are one order lower than that of natural graphite ($10^{-2}$ Ω m$^2$), indicating slower reaction kinetics of natural graphite probably due to its high orientation of Li$^+$ intercalation.

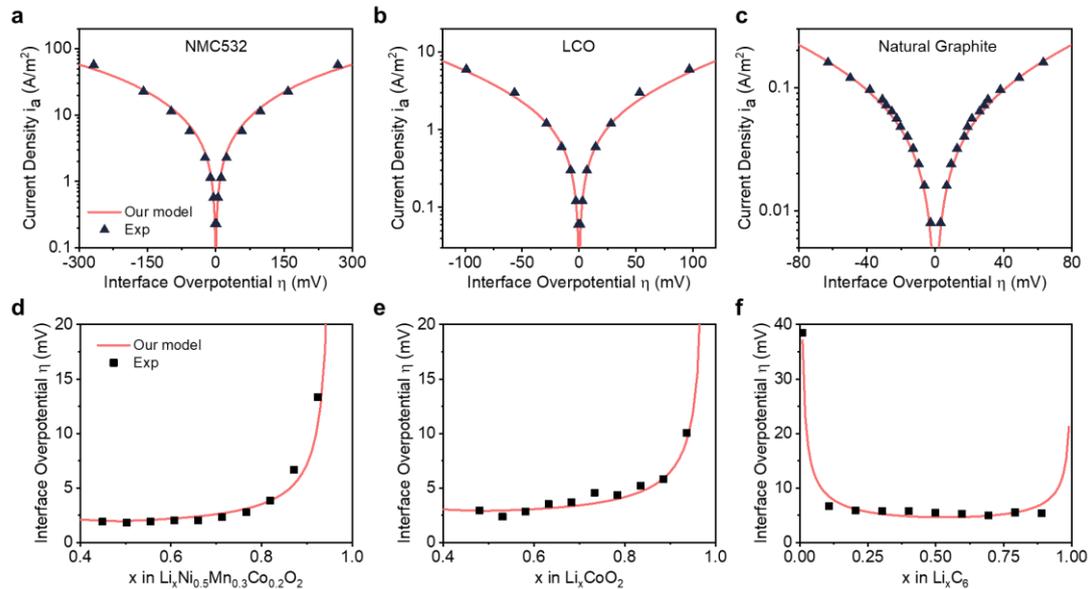

**Fig. 3 Model validation in interface reaction kinetics.** a-c. Current density-overpotential relation of the interface reaction kinetics compared with the interface ion-intercalation model. d-f. Dependence of interface reaction overpotential on solid-state Li$^+$ concentration of active particle surfaces compared with the interface ion-intercalation model. a,d. NMC532, b,e. LCO, c,f. Natural graphite.

**Table 1** Kinetics parameters of the interface ion-intercalation model of various materials.

| Parameter | $k$ (m$^{2.5}$ mol$^{-0.5}$ s$^{-1}$) | $\rho$ ($\Omega$ m$^2$) |
|---|---|---|
| NMC532 | $6\times10^{-11}$ | $1.3\times10^{-3}$ |
| LCO | $1.5\times10^{-11}$ | $1.5\times10^{-3}$ |
| Natural graphite | $2\times10^{-12}$ | $1.0\times10^{-2}$ |

**4.2 Modeling the reported ultrahigh-rate behaviors of single particles**

The developed interface ion-intercalation model is used to simulate the ultrahigh-rate behaviors of single-particles reported in the literature (the detailed mathematical modeling is given in **Eqn. S1-S8**), which is beyond the predicting capability of the Butler-Volmer model. As shown in **Fig. 4**, the charging/discharging profiles simulated based on the Butler-Volmer model and our interface ion-intercalation model are compared with the testing data from Ref. [19] (LCO) and Ref. [42] (mesocarbon microbead, MCMB), respectively, in which the discharging rate of LCO is from 13C (2 nA) to 1274C and the charging rate of MCMB is from 4.5C (9 nA) to 1485C. At lower C-rates, the simulation results based on the Butler-Volmer model fit the testing data well due to the slight polarization (**Fig. 4**a,c); however, there is a typical discrepancy between the simulation profiles and the testing results at higher C-rates, particularly over 300C for LCO and 500C for MCMB. The simulated voltage profiles remain obvious plateau characteristics similar to the equilibrium potential, while the plateau of actual voltage profiles gradually disappears with the increasing C-rate because the interface reaction overpotential increases dramatically as the charge/discharge proceeds (**Fig. S3**). Besides, there is also a non-negligible deviation from the practical capacity under ultrahigh-rate conditions; in contrast, the interface ion-intercalation model successfully reproduces the voltage behaviors of single particles at ultrahigh C-rates (**Fig. 4**b,d). Furthermore, it shows good applicability to various materials reported so far (**Fig. 5**) in addition to LCO and MCMB, including LiFePO$_4$ (LFP) [20], Li(Ni$_{0.8}$Co$_{0.15}$Al$_{0.05}$)O$_2$ (NCA) [27], Soft Carbon [43], SiOC [21], LiNi$_{0.5}$Mn$_{1.5}$O$_4$ (LNMO) [44], Li$_4$Ti$_5$O$_{12}$ (LTO) [22].

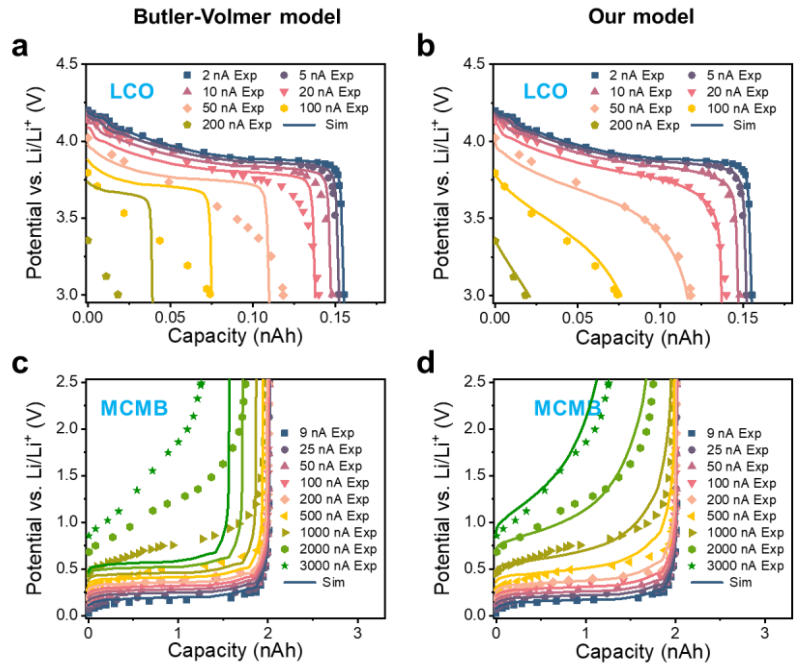

**Fig. 4 Model comparison in charging/discharging simulation of single particles.** Comparison of the simulation results based on the Butler-Volmer model (a, c) and the interface ion-intercalation model developed in this work (b, d). The experimental data of LCO (a,b) and MCMB (c,d) comes from Ref. [19] and Ref. [42], respectively.

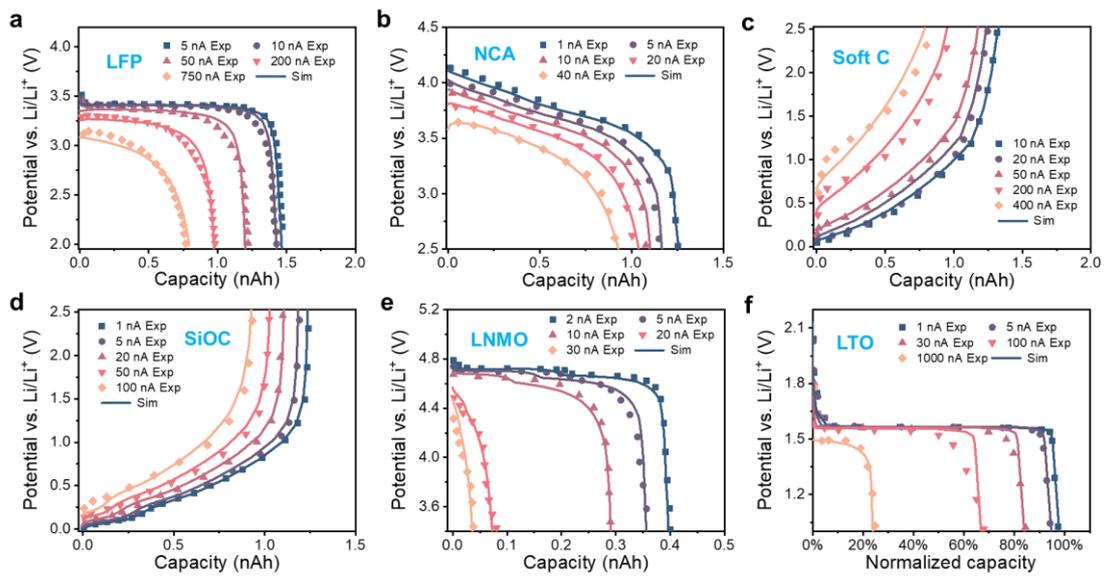

**Fig. 5 Charging/discharging simulations of various active materials.** a. LFP [20], b. NCA [27], c. Soft Carbon [43], d. SiOC [21], e. LNMO [44], f. LTO [22].

The estimated charge transfer constants $k$ of our model are one or two orders higher than that of the Butler-Volmer model, indicating faster reaction kinetics without such an ion-intercalation step (**Table S2**). The Li$^+$ (de)intercalation constants $\rho$ mainly range from 10$^{-4}$ Ω m$^2$ to 10$^{-3}$ Ω m$^2$ with a dimension of film resistance, among which the minimum value (4×10$^{-6}$ Ω m$^2$) belonging to LTO implicates ultra-fast Li$^+$ intercalation kinetics, in agreement with the fast-charging capability of LTO material [45,46]. Moreover, the temperature and aging dependence of the charge transfer constants $k$ and the Li$^+$ (de)intercalation constants $\rho$ revealed by the voltage data further verifies the reasonability of the interface ion-intercalation model. In the case of LTO [22], the temperature dependence of both parameters can be well described by the Arrhenius relation (**Fig. S3**a, **Eqn. S9**, and **Eqn. S10**), deriving the activation energies of 21.3 kJ/mol for $k$ (in agreement with the values reported in Ref. [47]) and 61.5 kJ/mol for $\rho^{-1}$. It validates the higher activation energy barrier in the solid-state environment than that in the liquid-state environment. According to the rate testing of the NCA particle after cycling 0, 400, and 800 cycles [27], $k$ exponentially decreases with the residual capacity while $\rho$ dramatically increases as the capacity fades (**Fig. S3b**), fitted by the empirical equations given in **Eqn. S12** and **Eqn. S13**. This aging dependence of $\rho$ on the host material further confirms that the aforementioned excess overpotential comes from the solid-state side of the interface, not the liquid-phase side.

### 4.3 Predicting composite electrode performance for high-power design

We further validated the interface ion-intercalation model at the electrode scale through porous electrode modeling (considering plenty of active particles stacking in the thickness direction), as demonstrated in **Fig. 6**a. Based on the model simulation, we revealed how the electrode rate performance declines from single particles to composite electrode structure, providing insights for high-power battery design. The rate-capacity relation of NMC532 electrodes with different areal capacities in a Li metal battery configuration was predicted (**Fig. 6**b). It shows an approximate exponential decline of

C-rate with the increasing areal capacity of electrodes from 0 (single particles) to 4.7 mAh/cm² (the commercial level). Enabled by overpotential calculations, the rate-capacity plot can be divided into three parts according to the dominant kinetic limiting processes, namely liquid-state Li⁺ transport limitation (blue), interface reaction limitation (red), and solid-state Li⁺ diffusion limitation (yellow). For single particles or thin electrodes with an areal capacity of 1 mAh/cm², the accessible capacity is determined by the solid-state Li⁺ diffusion at low C-rates (below 5C) while by the interface reaction process at medium and high C-rates. As demonstrated in **Fig. 6**c, interface reaction contributes the largest proportion to the total overpotential between the equilibrium potential ($E_{eq}$) and the cell voltage ($V_{cell}$) at the end of discharge at 20C (only half of the electrode capacity is accessible at such a high C-rate). **Fig. 6**d shows the critical case with an areal capacity of 2 mAh/cm² that the accessible capacity is together controlled by interface reaction and liquid-state Li⁺ transport at 12C. For thicker electrodes with an areal capacity above 2 mAh/cm², solid-state Li⁺ diffusion and interface reaction are still the dominant kinetic limitations at low and medium C-rates, respectively, while at higher C-rates, the electrodes will jump into the kinetic limitation region of liquid-state Li⁺ transport in the electrolyte solution, leading to a dramatic decline of accessible capacity. As shown in **Fig. 6**e, the overpotential component of liquid-state Li⁺ transport can reach as large as 0.6 V at 8C for the electrode with an areal capacity of 3 mAh/cm², limiting the operating at higher C-rates. Therefore, this rate-capacity plot explicitly provides the principles for designing high-power batteries with specific areal capacities.

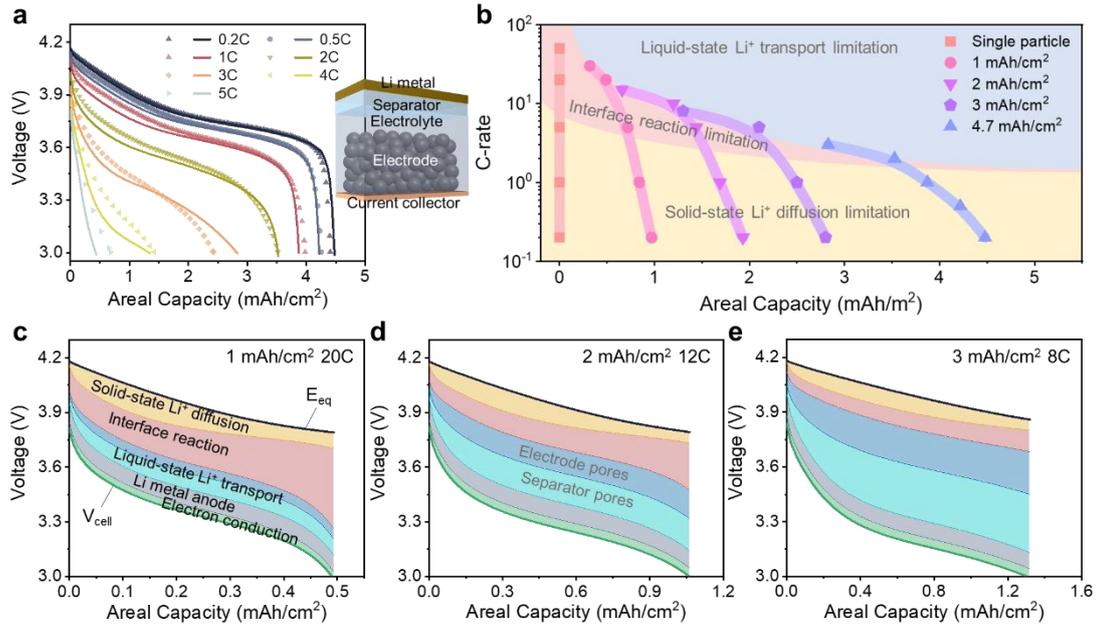

**Fig. 6 Model prediction of electrode-scale performance for high-power design.** a. Model validation of rate-performance prediction at the electrode scale. Solid lines: simulated profiles, dots: measured data. The fabricated NMC532 composite electrode with an areal capacity of 4.7 mAh/cm$^2$ was assembled in a Li metal battery configuration for rate-discharging testing. b. Rate-capacity plot of composite electrodes divided by three kinetic limitation regions according to overpotential component analysis. c-e. Components of the total overpotential between the equilibrium potential and the cell voltage in the cases of 1 mAh/cm$^2$ at 20C (c), 2 mAh/cm$^2$ at 12C (d), and 3 mAh/cm$^2$ at 8C (e), respectively. Each colored area denotes the corresponding overpotential of the labeled kinetic process. For example, the blue area corresponds to the overpotential of liquid-state Li$^+$ transport, including the dark part (from electrode pores) and the light part (from separator pores).

# 5 Conclusion

The interface reaction process of ion-intercalation active electrodes was decoupled from the reaction-diffusion coupled kinetics by time-resolved potential measurements, revealing that the classical Butler-Volmer equation deviates from the actual relation between current density, overpotential, and Li$^+$ concentration. By considering the ion-intercalation driving force in the interface reaction of ion-intercalation active materials, an interface ion-intercalation model was developed. Based on this model, charging/discharging behaviors of various active materials can be accurately predicted at both single-particle and electrode scales. We further revealed how the electrode rate performance declines from single particles to composite electrode structure and the kinetic limiting processes of electrodes with specific areal capacities for designing high-power batteries.

## Experimental section

**Electrode fabrication.** The electrode slurry was prepared by mixing NMC532 (MTI), carbon black (Super-P, MTI), and polyvinylidene fluoride (PVDF, Arkema) in N-methyl-2-pyrrolidone (NMP, Sinopharm Chemical Reagent Inc.) at a rate of 500 r/min for 6.5 h. The weight ratio of SLPEs was 3:0.5:1:20 (NMC532/Super-P/PVDF/NMP), and that of composite electrodes was 20:1:1:20. The slurry was coated on an aluminum foil by a blade coater with a gap of 40 μm, after which the wet electrode was dried at 100 °C by infrared radiation for 3 h. The dried electrodes were then calendered to a porosity of about 0.4 by a roller press. Before assembling the electrodes into cells, they were baked at 80 °C in a vacuum overnight.

**Cell assembly and measurements.** The electrodes were assembled into 2032-type coin cells with a Li-metal anode and a commercial separator (Celgard 2400) soaked in an electrolyte solution (DoDoChem, 1M $LiPF_6$ and EC/EMC/DMC at a volume ratio of 1:1:1 with 1.0% VC). Charging/discharging tests were performed on a battery cycler (LANHE, M340A). In the formation step, a constant-current constant-voltage (CC-CV) charging and CC discharging (CC-CV/CC) protocol was applied with a voltage window of 3-4.2 V. Symmetrical pulse charging/discharging were used to obtain by Tafel plots, in which each pulse lasted for 10 s with a high-speed sampling (1 kHz) after 30 min of resting. Piecewise discharging, in which cells were first charged to 4.2 V with a CC-CV protocol and then repeatedly rested for 1h and discharged at C/5 for 30 min until the potential dropped to 3 V, was conducted to construct the concentration-dependent overpotentials The high-speed sampling at 1 kHz was set during the initial 10 s of every discharge stages. All the tests above were carried out in a temperature-controlled room at a temperature of 25 °C.

**Electrochemical modeling and simulations.** The mathematical model of single-particle charging/discharging was given in Eqn. S1-S8. The composite electrode-based cells were modeled in the framework of porous electrode theory with the developed interface ion-intercalation model as the reaction kinetic equation. The detailed modeling description and the overpotential calculation method can be found in Ref. [34].

# Acknowledgment

This work was supported by the National Natural Science Foundation of China (Grant No. 52175317) and the Fundamental Research Funds for the Central Universities (No. 2021yjsCXCY023).

# Conflict of Interest

The authors declare no conflict of interest.

Supplementary information for

# Decoupled measurement and modeling of interface reaction kinetics of ion-intercalation battery electrodes

Ruoyu Xiong [a], Mengyuan Zhou [a], Longhui Li [a], Jia Xu [a], Maoyuan Li [a], Bo yan [b], Dequn Li [a], Yun Zhang [a, *], Huamin Zhou [a, **]

* Corresponding author: marblezy@hust.edu.cn, Tel: 86-27-87543492

** Corresponding author: hmzhou@hust.edu.cn, Tel: 86-27-87543492

a. State Key Laboratory of Material Processing and Die & Mould Technology, School of Materials Science and Engineering, Huazhong University of Science and Technology, Wuhan 430074, China

b. School of Materials Science and Engineering, Shanghai Jiao Tong University, Shanghai, 200030, China

# Mathematical modeling of single-particle charging/discharging

The configuration of the single-particle measurements is illustrated in **Fig. S1**a: a single active particle is attached to a metal filament (the current collector) soaked with an electrolyte solution, and the counter electrode is a Li metal sheet with a large area relative to the active particle [1,2]. The corresponding electrochemical structure can be simplified for modeling, as sketched in **Fig. S1**b. The active particles used in the single-particle measurements all had high sphericity, so a spherical assumption is justified. Though there are a few pores in some materials, like NMC and NCA, the total porosity is too low and hence neglected in model geometry [3]. The active particle is assumed to be equipotential for the small size of electron conduction. The electric resistance of the metallic current collector and the contact resistance between the active particle and the current collector are both ignored [4,5]. Outside the active particle, only the diffusion layer in the electrolyte solution is considered for the concentration polarization. Between the active particle and electrolyte, there exists a particle-electrolyte interface, of which the thickness is ignored. For the last part, the Li metal electrode is treated as an outer boundary of the diffusion layer. The surface overpotential of the Li metal electrode is neglected here for the very low current density. In addition, the model is assumed to be isothermal since the large amount of electrolyte solution surrounding the active particles and the metallic current collector can quickly take heat away to maintain a relatively stable temperature. The background current of the single-particle system is not taken into account for it was minimized in experiments [6].

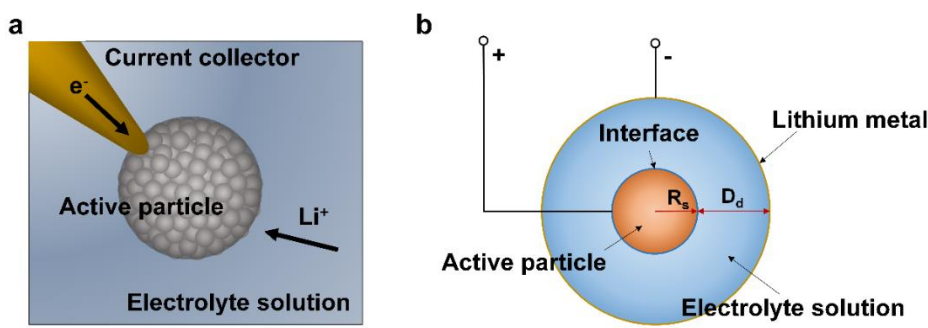

**Fig. S1** Schematic of the single-particle measurements (a) and the responding electrochemical

modeling (b).

In the active particle, the solid-phase diffusion process of Li$^+$ (polarons) is generally described by Fick's second law

$$\frac{\partial c_s}{\partial t} = \nabla \cdot (D_s \nabla c_{Li}) \tag{S1}$$

where $c_s$ is the Li$^+$ concentration in the solid phase (active particles), $D_s$ is the effective diffusion coefficient. Based on the concentrated solution theory, the transport process of Li$^+$ in the diffusion layer (electrolyte) is governed by

$$\frac{\partial c_l}{\partial t} = \nabla \cdot (D_l \nabla c_l) - \nabla \frac{t_+ \mathbf{i}_l}{F} \tag{S2}$$

where $c_l$ is the Li$^+$ concentration of the electrolyte solution, $D_l$ the diffusion coefficient, $t_+$ the Li$^+$ transference number, $\mathbf{i}_l$ the current density in the liquid phase, and $F$ the Faraday constant. The charge transport in the electrolyte solution can be expressed as

$$0 = \nabla \cdot \left( -\kappa \nabla \phi_l + \frac{2RT\kappa(1-t_+)}{F} \left(1 + \frac{\partial \ln f_\pm}{\partial \ln c_l}\right) \nabla \ln c_l \right) \tag{S3}$$

where $\phi_l$ is the electrostatic potential of the electrolyte solution, $\kappa$ the electrolyte conductivity, $f_\pm$ the activity coefficient of the electrolyte, and $R$, $T$ are the universal gas constant and the temperature, respectively. The overpotential of the particle-electrolyte interface is defined as:

$$\eta = \phi_s - \phi_l - E_{eq} \tag{S4}$$

where $E_{eq}$ denotes the thermodynamic equilibrium potential of the intercalation compounds, and $\phi_s$ is the electrostatic potential of the active particle.

The boundary conditions of the active particles are

$$-D_s \frac{\partial c_s}{\partial r}\bigg|_{r=0} = 0, \quad -D_s \frac{\partial c_s}{\partial r}\bigg|_{r=R_s} = -\frac{i_a}{F} \tag{S5}$$

The boundary conditions of the mass transport in the electrolyte solution are

$$-D_l \left.\frac{\partial c_l}{\partial r}\right|_{r=R_s} = -\frac{i_a}{F}, \quad -D_l \left.\frac{\partial c_l}{\partial r}\right|_{r=R_s+D_d} = -\frac{i_a}{F} \cdot \frac{R_s^2}{(R_s+D_d)^2} \tag{S6}$$

and the boundary conditions of the charge transport are

$$-\kappa \left.\frac{\partial \phi_l}{\partial r}\right|_{r=R_s} = -i_a, \quad -\kappa \left.\frac{\partial \phi_l}{\partial r}\right|_{r=R_s+D_d} = -i_a \cdot \frac{R_s^2}{(R_s+D_d)^2} \tag{S7}$$

The electrostatic potential of the Li counter electrode is set as

$$\phi_{Li} = 0 \tag{S8}$$

The model was numerically solved in COMSOL Multiphysics 5.4 with the parameter information extracted from the corresponding literature on single-particle measurements, as listed in **Table S1**. The particle radius $R_s$, ambient temperature $T$ were from the corresponding literature of the single-particle measurements. The maximum Li$^+$ concentration $c_{s,max}$ was calculated in terms of the theoretical capacity of the active materials. The Li$^+$ diffusion coefficient $D_{l,0}$, ion conductivity $\kappa_0$, and partial active coefficient $\frac{\partial f_\pm}{\partial c_l}$ of the electrolyte were from Ref. [7-9].

**Table S1** Model parameters of various active materials.

| Material | LCO | MCMB | LFP | Soft carbon | SiOC | LTO | LNMO | NCA |
|---|---|---|---|---|---|---|---|---|
| $R_s$ (μm) | 3.8 | 9 | 8.7 | 8.9 | 6.7 | 9 | 6.8 | 8.7 |
| T (°C) | 25 | 25 | 30 | 25 | 25 | 30 | 25 | 25 |
| $c_{Li,max}$ (mol/m$^3$) | 56250 | 25000 | 22800 | 19000 | 40300 | 22400 | 24100 | 20900 |
| $D_{l,0}$ (m$^2$/s) | 1.2×10$^{-10}$ (at 1M, 25 °C) | | | | | | | |
| $\kappa_0$ (S/m) | 0.5 (at 1M, 25 °C) | | | | | | | |
| $t_+$ | 0.363 (at 1M, 25 °C) | | | | | | | |
| $\partial f_\pm/\partial c_l$ | 0.6 (at 1M, 25 °C) | | | | | | | |

Note: the electrolyte parameters are all concentration and temperature-dependent.

# Estimated kinetics parameters of the interface ion-intercalation model

Table S2 Kinetics parameters of the interface ion-intercalation model of various active materials.

| Parameter | $k$ (m$^{2.5}$ mol$^{-0.5}$ s$^{-1}$) | $\rho$ (Ω m$^2$) |
|---|---|---|
| LCO | $3\times10^{-9}$ | $1.8\times10^{-4}$ |
| MCMB | $1\times10^{-10}$ | $1.2\times10^{-4}$ |
| LFP | $9\times10^{-10}$ | $1.3\times10^{-4}$ |
| Soft Carbon | $3\times10^{-10}$ | $1\times10^{-3}$ |
| SiOC | $1\times10^{-9}$ | $2\times10^{-4}$ |
| LTO | $2\times10^{-8}$ | $4\times10^{-6}$ |
| LNMO | $1\times10^{-8}$ | $2.0\times10^{-3}$ |
| NCA | $1\times10^{-10}$ | $1.5\times10^{-3}$ |

# Rate-limiting process of single particles

Taking LCO as an example, **Fig. S2**a shows the concentration polarization in the active particle and the electrolyte solution at the end of discharging at 127C (dashed lines) and 1274C (solid lines). There is a slight concentration gradient of $Li^+$ in the electrolyte solution in both cases, and it causes a negligible overpotential (**Fig. S2**b) owing to the relatively high ion conductivity, high diffusion coefficient, and sufficient electrolyte. Therefore, the length of the diffusion layer assumed in the model (30 μm) is hence demonstrated to be reasonable. For the $Li^+$ concentration within the active particle, it displays a more apparent gradient and 1274C than that at 127C, whereas the concentration overpotential at 127C is much higher due to the nearly saturated Li content at the particle surface. Thus, the $Li^+$ diffusion process in the solid phase is revealed to be the bottleneck of the system kinetics below 150C (**Fig. S2**b). Above this C-rate, the rate-limiting process of the single particles shifts to the interface reaction process, more specifically, the $Li^+$ (de)intercalation step. As a result of the large interface reaction overpotential, the discharge rapidly ends before the solid-state $Li^+$ concentration reaches a higher level at the particle surface, leading to a lower material utilization. Nevertheless, fast solid-state diffusion kinetics also helps reduce the interface reaction overpotential because of the concentration dependence of $Li^+$ intercalation kinetics.

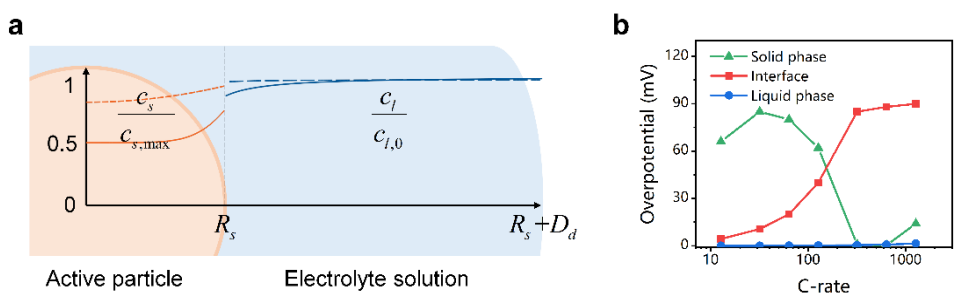

**Fig. S2** a. $Li^+$ concentration distribution in the solid/liquid phase at the end of discharge at 127C (dashed lines) and 1274C (solid lines); b. Components of the total overpotential at the end of discharge.

# Temperature and aging dependence of the kinetic parameters

The reaction constant $k$ was fitted by the Arrhenius equation

$$k = k_0 \exp\left[\frac{E_a}{R}\left(\frac{1}{T_{ref}} - \frac{1}{T}\right)\right] \quad (S9)$$

where $k_0 = 2\times10^{-8}$ m$^{2.5}$ mol$^{-0.5}$ s$^{-1}$, $E_a = 21.3$ kJ/mol, and $T_{ref}$ is the reference temperature. The insertion constant $\rho$ was fitted by the Arrhenius equation with a reciprocal form

$$\frac{1}{\rho} = \frac{1}{\rho_0} \exp\left[\frac{E_a}{R}\left(\frac{1}{T_{ref}} - \frac{1}{T}\right)\right] \quad (S10)$$

where $\rho_0 = 4\times10^{-6}$ Ω m$^2$, $E_a = 61.5$ kJ/mol. The aging state of the active particles was defined as the percentage of the residual capacity (also used as the definition of SOH sometimes)

$$C_r = \frac{C_{ini} - C}{C_{ini}} \times 100\% \quad (S11)$$

Based on the estimated values of $k$ after 0, 400, and 800 cycles, the dependence on the aging state can be fitted by an empirical equation

$$k = k_{ini} \exp\left[-17.05(1-C_r)\right] \quad (S12)$$

where $k_{ini} = 5\times10^{-10}$ m$^{2.5}$ mol$^{-0.5}$ s$^{-1}$ and the insertion constant $\rho$ can be fitted by

$$\rho = \rho_{ini} + 1.2(1-C_r)^{3.5} \quad (S13)$$

where $\rho_{ini} = 1.4\times10^{-3}$ Ω m$^2$.

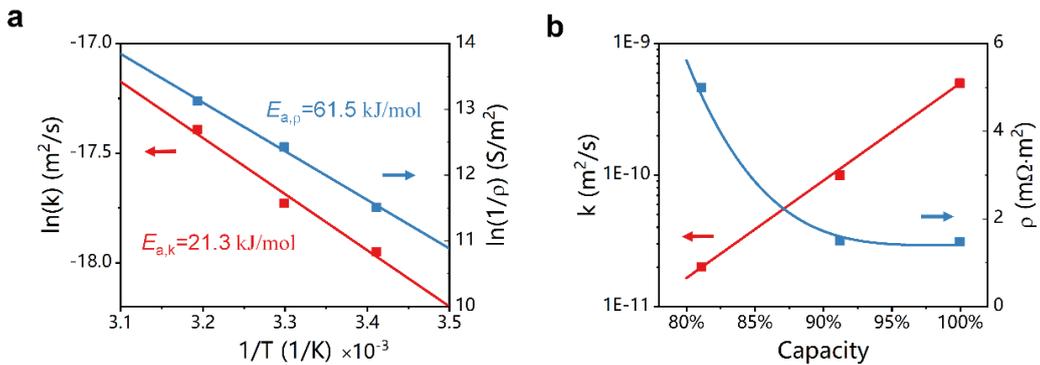

**Fig. S3** Dependence of the kinetic constants in our interface ion-intercalation model on temperature (a) and residual capacity (b). The data derives from LTO [4] and NCA [10], respectively.

# List of terms

| | |
|---|---|
| $i_a$ | current density |
| $\eta$ | overpotential |
| $\alpha$ | charge transfer coefficient |
| $F$ | Faraday constant |
| $R$ | universal gas constant |
| $T$ | temperature |
| $i_0$ | exchange current density |
| $\rho$ | Li$^+$ (de)intercalation constant |
| $k$ | charge transfer constant |
| $c$ | species concentration |
| $D$ | diffusion coefficient |
| $t_+$ | Li$^+$ transference number |
| $\phi$ | electrostatic potential |
| $f_\pm$ | activity coefficient |
| $\kappa$ | electrolyte conductivity |
| $E_{eq}$ | thermodynamic equilibrium potential |
| $R_s$ | particle radius |
| $D_d$ | the thickness of the diffusion layer |
| $R_{in}$ | the Li$^+$ (de)intercalation resistance in the context |

*Subscript*

| | |
|---|---|
| $s$ | solid phase |
| $l$ | liquid phase |
| $surf$ | active particle surface |
| $in$ | Li$^+$ (de)intercalation |
| $Li$ | Li metal anode |
| $site$ | active site |